\documentclass[a4paper,11pt]{article}
\pdfoutput=1 

\usepackage{jinstpub} 

\usepackage{lineno}
\title{\boldmath Fabrication and performance of AC-coupled LGADs}

\author[1]{Gabriele Giacomini,\note{Corresponding author.}}
\author{Wei Chen,}
\author{Gabriele D'Amen,}
\author{Alessandro Tricoli}

\affiliation{Brookhaven National Laboratory,\\Upton, 11973, NY, USA}

\emailAdd{giacomini@bnl.gov}

\abstract{Detectors that can simultaneously provide fine time and spatial resolution have attracted wide-spread interest for applications in several fields such as high-energy and nuclear physics as well as in low-energy electron detection, photon science, photonics and imaging.  Low-Gain Avalanche Diodes (LGADs), being fabricated on thin silicon substrates and featuring a charge gain of up to 100, exhibit excellent timing performance. Since pads much larger than the substrate thickness are necessary to achieve a spatially uniform multiplication, a fine pad pixelation is difficult. To overcome this limitation, the AC-coupled LGAD approach was introduced. In this type of device, metal electrodes are placed over an insulator at a fine pitch, and signals are capacitively induced on these electrodes. At Brookhaven National Laboratory, we have designed and fabricated prototypes of AC-coupled LGAD sensors. The performance of small test structures with different particle beams from radioactive sources are shown. 
}

\keywords{Solid state detectors; Timing detectors; Charge transport and multiplication in solid media; Detector modelling and simulations II (electric fields, charge transport, multiplication and induction, pulse formation, electron emission, etc)}

\begin{document}
\maketitle
\flushbottom

\section{Introduction}
Low-Gain Avalanche Diodes (LGADs)~\cite{Fernandez-Martinez:2015bda} are a new class of silicon sensors that have recently attracted the attention of the high-energy physics community thanks to their fast-timing properties~\cite{Sadrozinski:2013nja, Collaboration:2623663,Collaboration:2296612}. They are fabricated on thin high-resistivity silicon $p$-type substrates (about 50-\textmu m thick) and are based on simple $p$-$n$ junctions: a large and shallow $n$++ implant covers a deep $p$+ layer. The integral of the boron doping concentration (i.e., the implantation dose) of this latter implant is typically a few units of $10^{12} ~{\rm cm^{-2}}$, and extends into the substrate for up to a few microns. Application of a bias voltage across this junction leads to a depletion of the $p$+ layer, creating an intense electric field: the dose and the doping profile of such an implant are engineered in such a way that the resulting electric field is above the threshold for electron impact ionization, and sufficiently low for any significant hole ionization to occur. The onset of a self-sustaining avalanche breakdown is thus excluded. Electrons traversing the device are then subject to a multiplication, in the range from 5 to 100, and current pulses at the terminals are mainly due to the drift of the multiplication holes through the substrate. The $p$+ layer is thus referred to as {\it gain} layer. The large signals and the thin substrates are instrumental for producing fast signals: a timing resolution of a few tens of picoseconds has been demonstrated~\cite{cartigliabeamtest}. 

In a silicon processing clean room, thin substrates can be worked only if they are attached to a thick supporting layer for handling: the thin substrate layer can be either an epitaxial layer, grown over a thick low-resistivity wafer or a thin wafer, wafer-bonded to another wafer and then thinned down at will. In both cases, the back acts as a uniform ohmic contact and all the processing takes place on the front side. Here, once the electrodes are patterned in shape of pads or pixels, 2D spatial information can be obtained. However, for a  uniform multiplication to occur along the entire surface of the sensor, the pad dimension must be far larger than the substrate thickness. For example, in the ATLAS High Granularity Timing Detector (HGTD)~\cite{Collaboration:2623663} the pixel size will be 1.3 mm $\times$ 1.3 mm, with an active substrate thickness of 35--50 \textmu m.  A further example is provided by the silicon microstrip sensors that implement the gain layer under the strip~\cite{WADA2019380}: such devices show multiplication only in a fraction of the area, i.e., in the center of the strip. A way to circumvent such problem can be the placement of a large uniform pad on the opposite side of the patterned electrodes~\cite{DALLABETTA2015154, PELLEGRINI201624}. However, as mentioned above, also in this case the wafer must be thick enough as to be processed in a standard clean-room (i.e., 200-300 \textmu m thick, depending on the wafer diameter) and thus the fast timing properties of the LGADs are compromised. A novel concept to preserve the fast timing and ensure a highly segmented detector is the AC-coupled LGAD (AC-LGAD), where signals are capacitively induced on metal pads placed on a thin dielectric layer, which is grown over the uniform $n$+ layer that extends over the entire active surface of the sensors. The gain layer will also be implanted uniformly over the entire active area of the sensor. For discussions of the AC-LGAD concept we refer to documents in refs.~\cite{firstAC,RSD, RD50, AIDAWP7}

At Brookhaven National Laboratory (BNL), we have successfully designed and fabricated standard (i.e., DC-coupled) LGADs and we have leveraged such expertise to develop and fabricate AC-LGADs. Details of the AC-LGAD concept are given in section 2, while differences in the layout and in the fabrication processes between the AC-LGAD and the standard LGAD are detailed in section 3. Static and functional measurements of the first production of this new class of devices at BNL are given in section 4.

\section{AC-LGAD concept}
\begin{figure}
\centering 
\includegraphics[width=.65\textwidth]{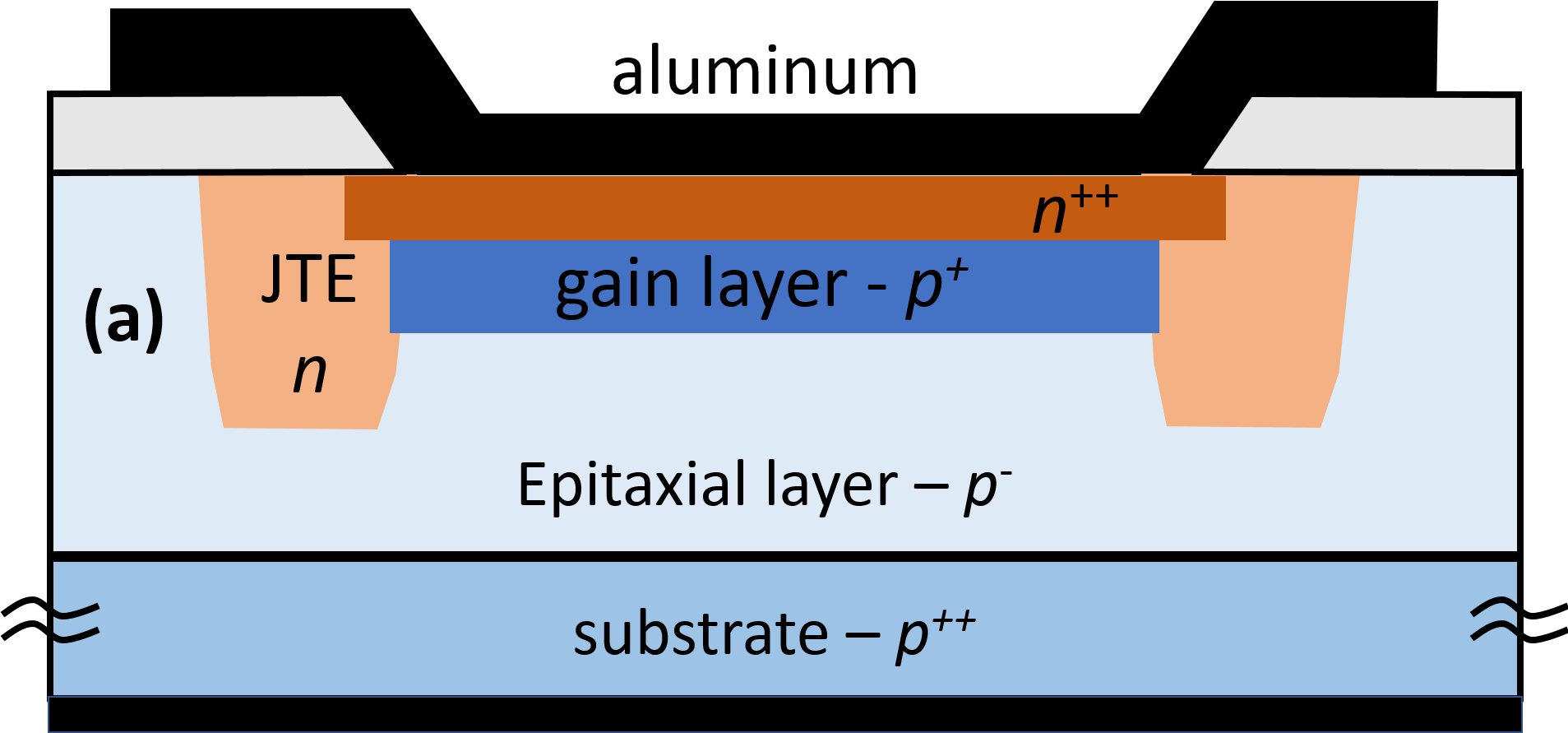}
\includegraphics[width=.65\textwidth]{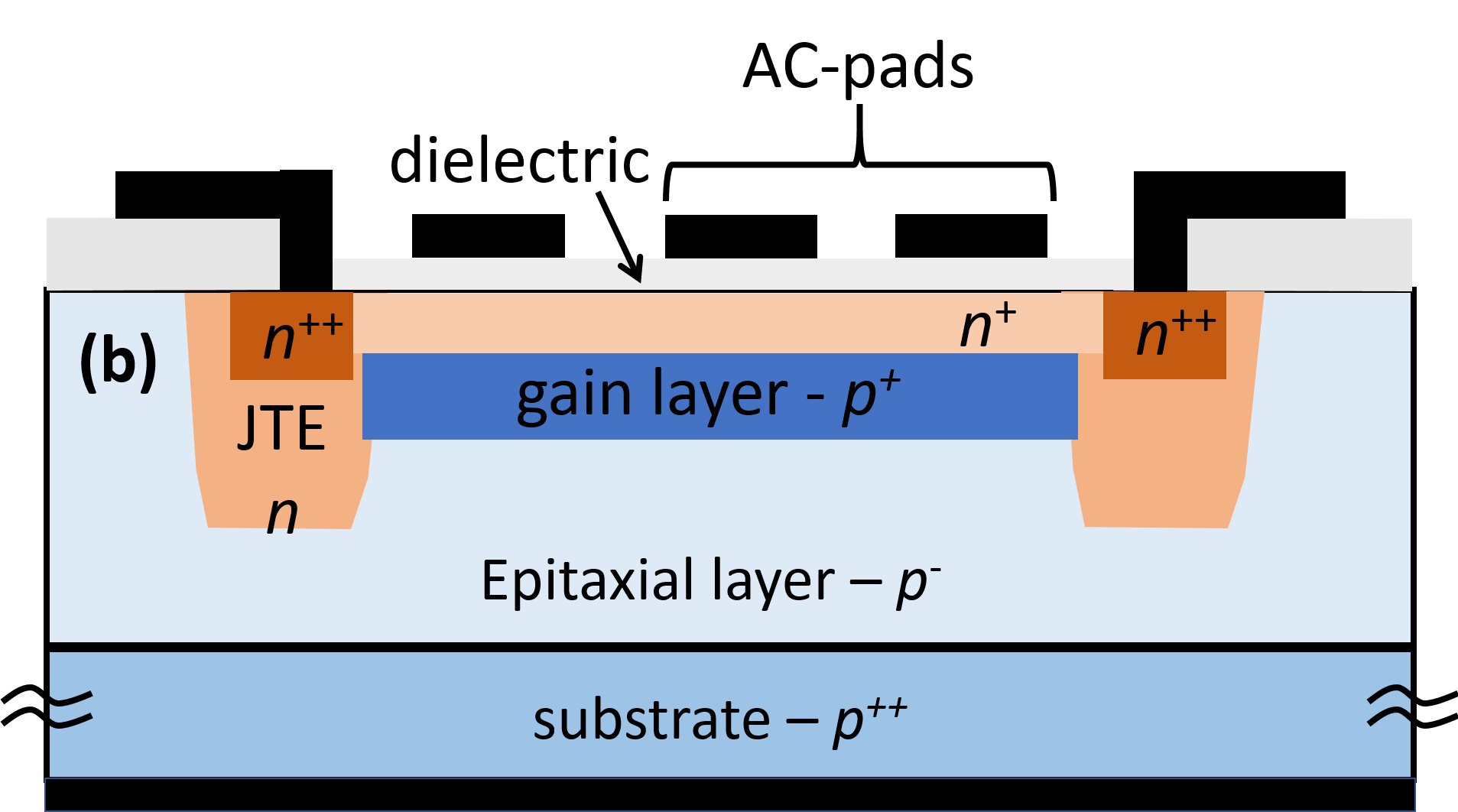}
\caption{\label{fig:sketch} (a) Sketch of a section of a single-pad  standard LGAD; (b) sketch of a section of a segmented AC-LGAD (not to scale). }
\end{figure}

A sketch of an AC-LGAD structure, as compared to a standard LGAD, is shown in figure~\ref{fig:sketch}.
In order to exhibit similar gains and in turn similar fast timing performance as the LGADs, AC-LGADs are fabricated on $p$--type substrates of the same thickness. Upon application of a bias voltage, uniformly implanted $n$++ and gain layers assure parallel electric field lines extending into the bulk, and in turn spatial uniformity of the gain. 

One of the main differences between AC-LGADs and standard LGADs is the replacement in AC-LGADs of the $n$++ layer by a much less doped $n$+ layer, see below for details. Moreover, the electrodes, which the read-out electronics is connected to, are metal pads  separated from the $n$+ layer by a thin insulator. Since the $n$+ layer is continuous, these metal electrodes over the insulator experience a low inter-pad resistance $R_{n}$, and this can cause two effects.
First, a potentially large thermal Johnson noise can be fed into the front--end electronics. This effect can be made negligible by reading out the electrodes with short shaping times: a fast read-out (sub-GHz bandwidth) is in any case beneficial to achieve optimal timing resolution.
Second, signal loss occurs if $R_{n} << (\omega C_{\rm AC})^{-1}$, where  $C_{\rm AC}$ is the effective capacitance of the metal pad towards the $n$+ layer. Considering a simple lumped model for a  complicated impedance network of sheet resistances and capacitances, $C_{\it AC}$ is given by the parallel plate approximation $C_{\rm AC}= \epsilon_{\rm diel} \cdot {\rm A} / {\rm d_{\rm diel}}$, where $A$ is the metal area and $d_{\rm diel}$ the thickness of the dielectric. 
It is clear that dielectrics as thin as possible are preferable: a limiting factor is that, for the same voltage difference between the $n$+ layer and the front-end electronics, electric fields inside the dielectric must not exceed the dielectric critical field $E_{\rm c}$, which causes dielectric rupture. For example, if the maximum voltage difference is 10 V and the critical oxide field is 1 MV/cm (for comparison, for a good thermal oxide, $E_{\rm c} \sim$ 6--9 MV/cm), the minimum dielectric thickness is 100 nm.  This consideration also sets a lower limit on the area of the pixel to ensure $C_{\rm AC}$ is large enough for a given $R_{n}$, and this may limit the size of the pixelation. By reducing the implantation dose of the $n$+ layer, i.e., making it as resistive as possible, we can increase $R_{n}$. The lower limit to the dose is set by the sum of the dose of the gain layer and the integral of the doping concentration of the substrate: this sum can be just shy of  $5 \cdot 10^{12} ~{\rm cm^{-2}}$. A larger value of the $n$+ dose can be foreseen to account for the over-depletion of the substrate, or in case large voltages are to be applied to achieve the target gain. 

In this design a highly-doping $n$++ implant is still present at the edge of the device, embedded into the Junction Termination Edge (JTE), and DC-connected to a voltage source for electron current draining.  
Electrons 
are first collected by the $n$+ layer and 
from it by an electrode contacting the $n$++ implant at the periphery of the device. The back of the device is uniform, and acts as an ohmic contact. At the edge of the active area of the device, the same termination used in LGADs is used for AC-LGADs: a JTE consisting of a deep diffused $n$+ implant prevents the development of high fringe electric fields in this regions, and thus the onset of early breakdowns~\cite{Fernandez-Martinez:2015bda}. Externally to the active area, a series of floating Guard Rings (GR) are included in the design.


Since  AC-coupled electrodes  do not collect  charge, the current induced on the AC-LGAD pads is bipolar with zero net integral, with a first peak accounting for the drift of the multiplication holes into the substrate and a second peak, of opposite polarity, to account for diffusion of the electrons within the $n$+ toward the $n$++ contact at the edge of the device. The latter charge movement will induce signals also in nearby pixels: this effect can be attenuated by making the $n$+ layer to float through a large value resistor $R_{\rm GND}$ connected between ground and the $n$++ contact. Electrons collected at the $n$+ layer will then discharge slowly to the $n$++ contact, with  a time constant given approximately by  $R_{\rm GND} \cdot C_{n+}$, where $C_{n+}$ is the capacitance of the $n$+ layer towards the rest of the world. This time constant is much larger than the read-out time, so that the opposite polarity pulse amplitude is negligible.


\section{Design and fabrication}
\begin{figure}
\centering 
\includegraphics[width=.8\textwidth]{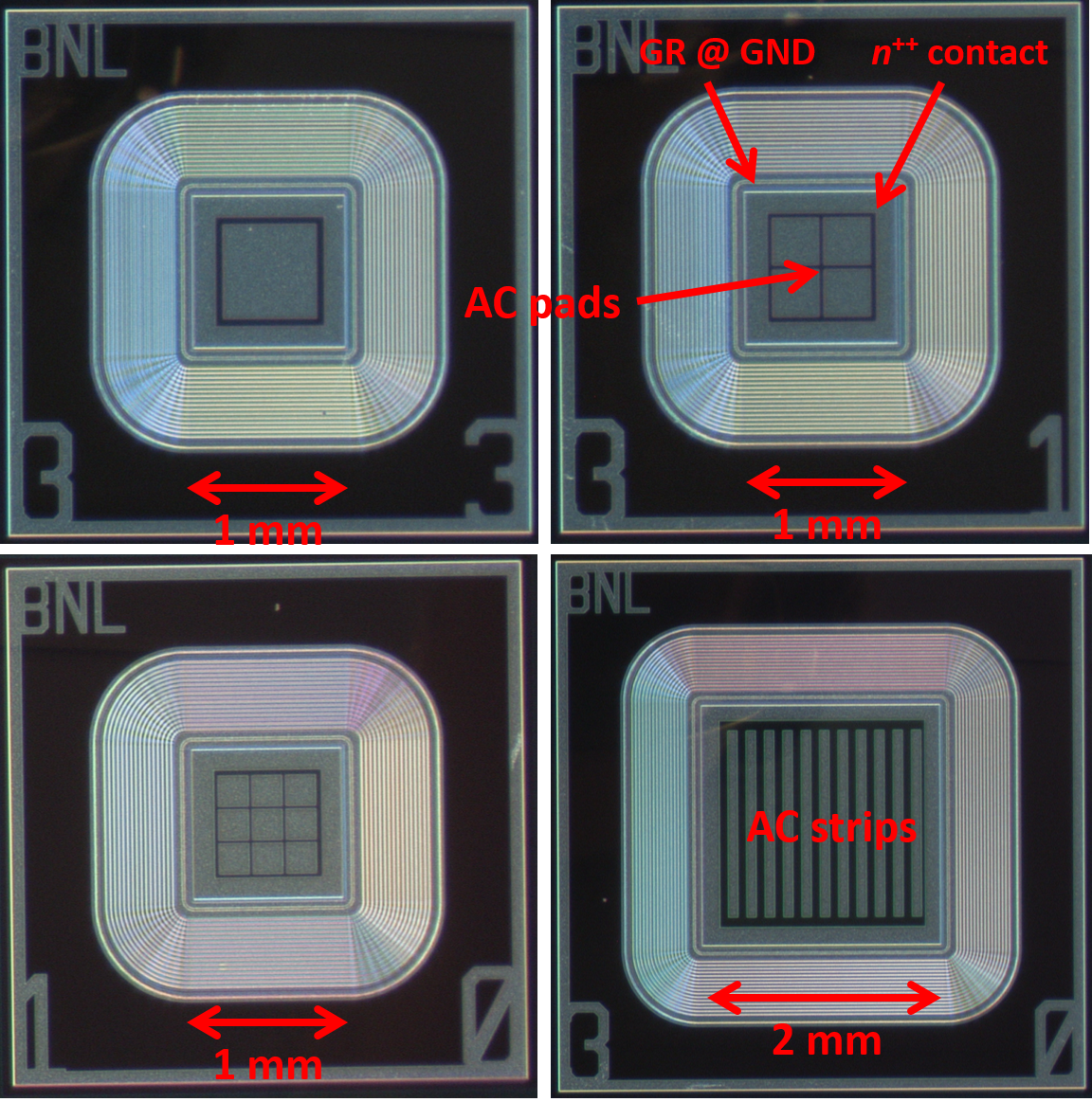}
\caption{\label{fig:pictures} Photographs of different AC-LGAD structures: (top-left) single AC-pad 600 \textmu m $\times$ 600 \textmu m active area, (top-right) 2 $\times$ 2 array with 300 \textmu m $\times$ 300 \textmu m pads at 330   \textmu m pitch, (bottom-left)  3 $\times$  3 array with 200 \textmu m $\times$  200 \textmu m pads at 220 \textmu m pitch, and (bottom-right) array of strips of 80 \textmu m width $\times$  1.5 mm length at 150 \textmu m pitch. The sizes of the active areas, the positions of the AC-coupled pads, the GR and $n$++ contacts are indicated with arrows.}
\end{figure}

For this first, proof of concept, production of AC-LGADs at BNL, photolithographic masks developed for previous LGAD fabrications at BNL were used, while three main changes took place: the $n$++ implant, the contact structures and  the metal layer. The following details the main changes.
\begin{itemize}
    \item The mask that in the LGAD production defines the $n$++ layer in AC-LGADs is used to define the resistive $n$+ layer; however this is 10-time less doped to increase its resistivity in AC-LGADs.
    \item A new mask was designed to define the $n$++ implantation at the edge of the device, embedded into the JTE (figure~\ref{fig:sketch}).
    \item A new mask 
    was designed to remove the contacts in the active areas, while contacting the $n$++ at the edge of the device.
    \item A new mask for the metal layer was designed. Several electrode arrays differing for pixel size and pitch were placed in the active area of small test structures (figure~\ref{fig:pictures}).
\end{itemize}

The process flow follows the one detailed in \cite{GIACOMINI201952} up to the step of $n$+ implantation. The $n$++ phosphorus implantation at the border of the  active area  follows, then a final activation of the dopant species (annealing) is performed. A 100-nm thickness of PECVD silicon nitride is deposited. Finally contacts are opened, aluminum is sputtered and then patterned.


It must be pointed out that the depletion of the $n$+ layer is not negligible, causing the gain layer to be effectively deeper from the junction than for standard LGADs. Impact ionization strongly depends on the distance that electrons travel in a high electric field, so that the dose of the gain layer must be accordingly adjusted (lowered) to avoid a premature breakdown. 

Figure~\ref{fig:pictures} shows some of the fabricated AC-LGAD structures, cut out from a wafer.

\section{Measurements of performance}
The performance of the AC-LGAD structures designed and fabricated at BNL is measured together with the one of LGADs and diodes that are fabricated on the same wafer as the AC-LGADs.  Measurements of static electrical performance, gain, induced signal, cross-talk between pixels, and timing are carried out and detailed in the following.

\subsection{Static measurements}
\begin{figure}
\centering 
\hspace{-5mm}
\includegraphics[width=.50\textwidth]{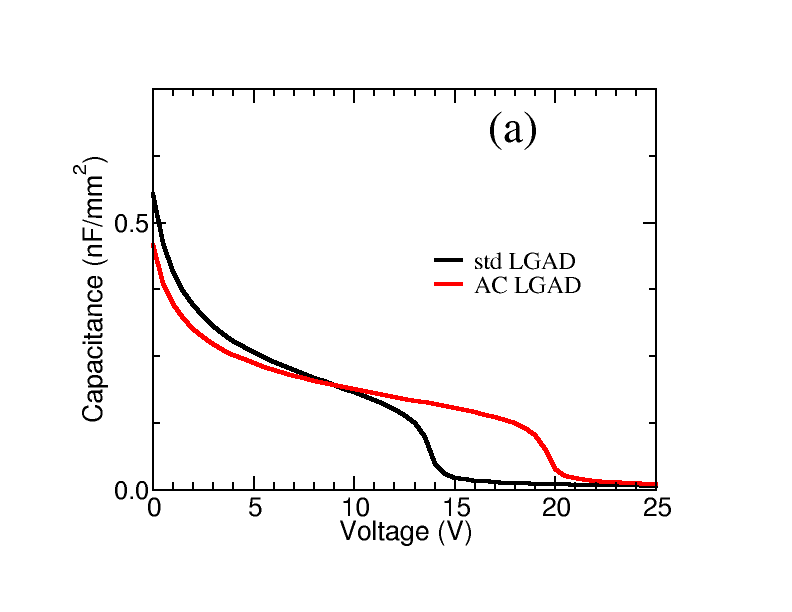}
\hspace{-10mm}
\includegraphics[width=.50\textwidth]{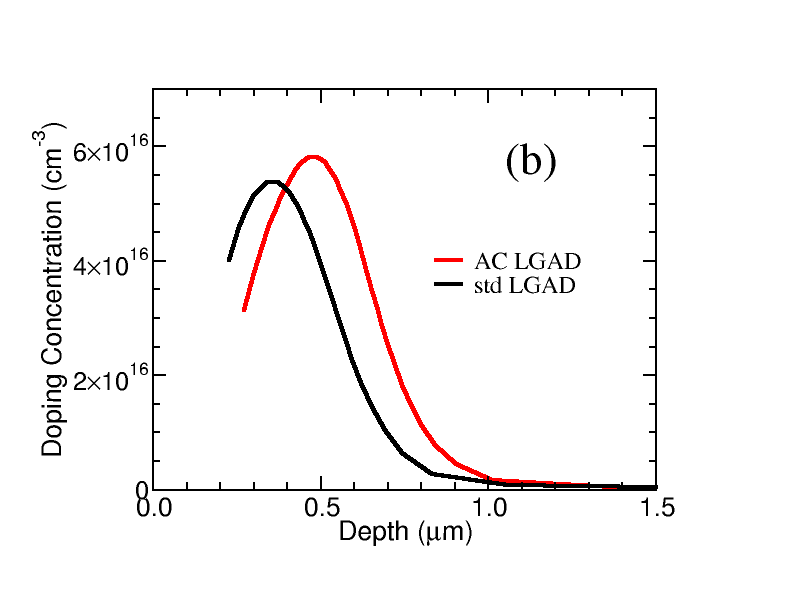}
\caption{\label{fig:profile} (a) Measurements of capacitance per unit area as a function of absolute value of the bias voltage in a standard LGAD and in an AC-LGAD; (b) the gain layer doping concentration as a function of depth, as extracted from the curves in (a), see formula in ref.~\cite{grove} , page 171. }
\end{figure}

 The main difference between the LGAD and AC-LGAD devices is the implantation dose of the phosphorus layer that makes the junction with the gain layer. In the LGAD case, a standard high dose is used, while in the AC-LGAD case a 10-time less doped implant is used to increase the resistance of such a layer. Figure~\ref{fig:profile} shows the measurements of the junction capacitance as a function of the bias voltage (C-V curves) for standard LGADs and AC-LGADs. The C-V curve measured on LGADs results compatible with those obtained in previous batches, as detailed in ref.~\cite{GIACOMINI201952}. The measurement of the C-V curve on AC-LGAD shows a {\it foot} at higher voltages, in absolute values, than for LGADs, i.e., at 20 V for the AC-LGAD with respect to 15 V for the LGAD. Such a result points towards a deeper gain layer, which is confirmed in the extracted doping concentration profile as a function of depth shown in figure~\ref{fig:profile} (b). Since the  zero depth of the extracted doping profile is at the $p$--$n$ junction, and the gain layer is the same for all structures on the wafer (i.e., LGADs and AC-LGADs),  these measurements suggest that a non-negligible depletion occurs in the $n$+ layer in AC-LGADs. As a result, higher gain is achieved in the AC-LGAD compared to the LGAD despite an identical set of gain layer implantation parameters.  As a consequence, the breakdown voltage of the AC-LGAD device is expected to be much lower, and in fact it is found at about $-$80 V whereas for the standard LGAD is measured to be at $-$300 V. Since in diodes (i.e., LGADs with unit gain)  the depletion voltage of the 50-\textmu m-thick epitaxial layer is measured to be $-$125 V, in LGADs fabricated on the same substrate the depletion is expected to be at $V_{\rm diode~depletion} + V_{\rm LGAD~foot}$. As expected, the depletion voltage measured on the standard LGADs is $-$140 V (i.e., $V_{\rm diode~depletion} + V_{\rm std LGAD~foot}$ = $-$125 $-$ 15 = $-$140 V), while on the AC-LGAD fabricated on the same wafer as the LGADs the depletion is expected to be at about $-$145 V (i.e., $V_{\rm diode~depletion} + V_{\rm AC-LGAD~foot}$). In the following measurements the AC-LGADs are therefore operated in under-depletion mode at $V_{\rm bias}$ = $-$80 V, close to their breakdown voltage.

\subsection{Gain}
To test the performance of the AC-LGAD as compared to  the standard LGAD, we mounted a sample of  each type of devices on a multi-channel board that consists of 16 input channels, each of which includes an independent Transimpedance Amplifier (TA). To characterize the TA, we also mounted a diode 
fabricated in the same wafer as the AC-LGAD, and we irradiated the diode with fluorescence X-rays emitted by a ${}^{65}$Tb target ($K_{\alpha} =$ 44 keV) excited by alpha particles from ${}^{241}$Am. By means of a 1 GHz scope, a number of the resulting signal waveforms were acquired 
and the average value of their integrals is measured to be $ 9.4 ~{\rm pV \cdot s}$, which provides the absolute calibration of the TA. Subsequently, in an AC-LGAD the $n$++ layer at the periphery of the device was DC-connected to a TA input and from the distribution of the waveform integrals from a $^{90}$Sr beta source
 an average  value of $ 204 ~{\rm pV \cdot s}$ is extracted. From the calibration and extracting from the C-V scan a depleted thickness of 32 \textmu m, a gain of about 90 is derived for a bias voltage of $-$80 V. 



\begin{figure}
\centering 
\includegraphics[width=.8\textwidth]{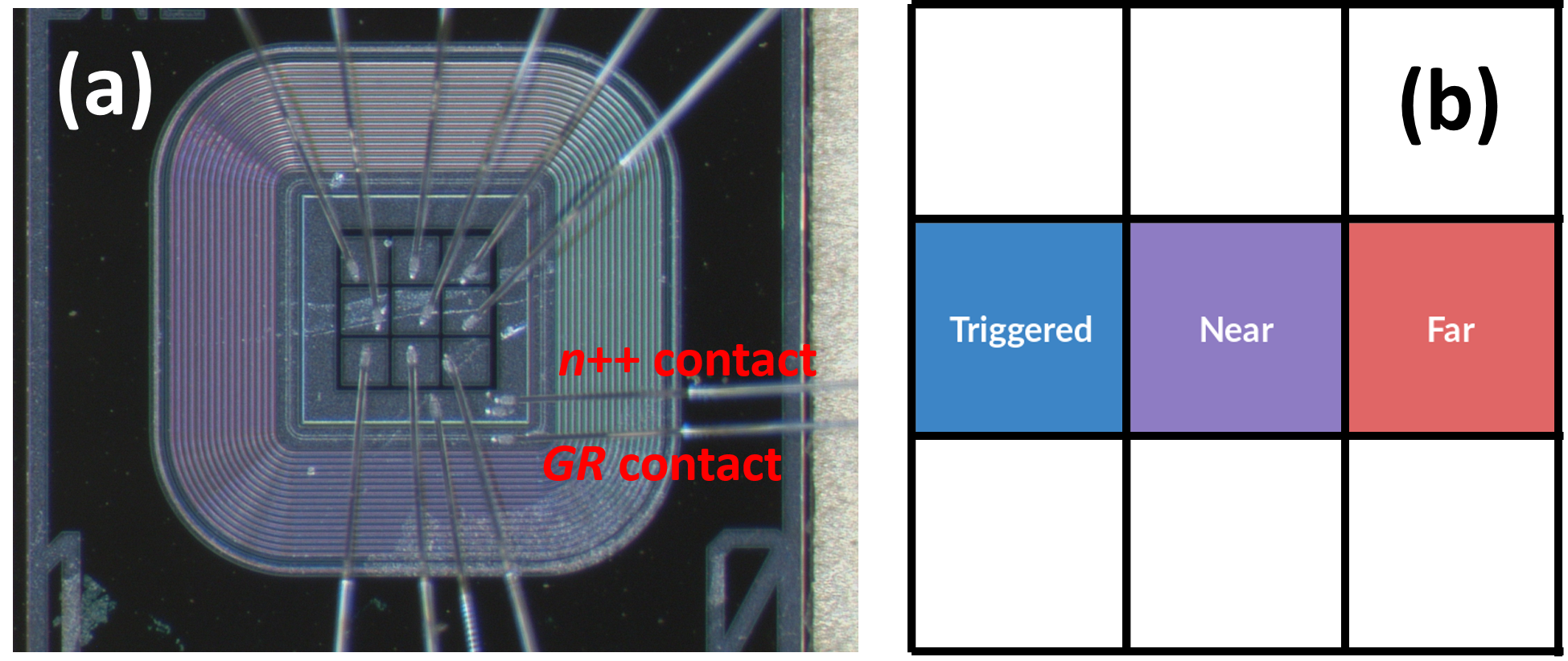}
\caption{\label{fig:sensors}  (a) Photo and (b) sketch of the 3 $\times$  3 array of AC-LGAD pixels with 200 \textmu m $\times$  200 \textmu m size and 220 \textmu m pitch used to study the cross-talk between pixels. The middle row of pixels is read out: the channel on the left, in blue colour, is used for triggering the event ({\it Triggered}), while its neighbouring channels, named as {\it Near}, in purple colour, and {\it Far}, in red colour, corresponding to the adjacent and next-to-adjacent pixels, respectively, are read out simultaneously together with the triggering channel.}
\end{figure}

\begin{figure}
\centering 
\includegraphics[width=.43\textwidth]{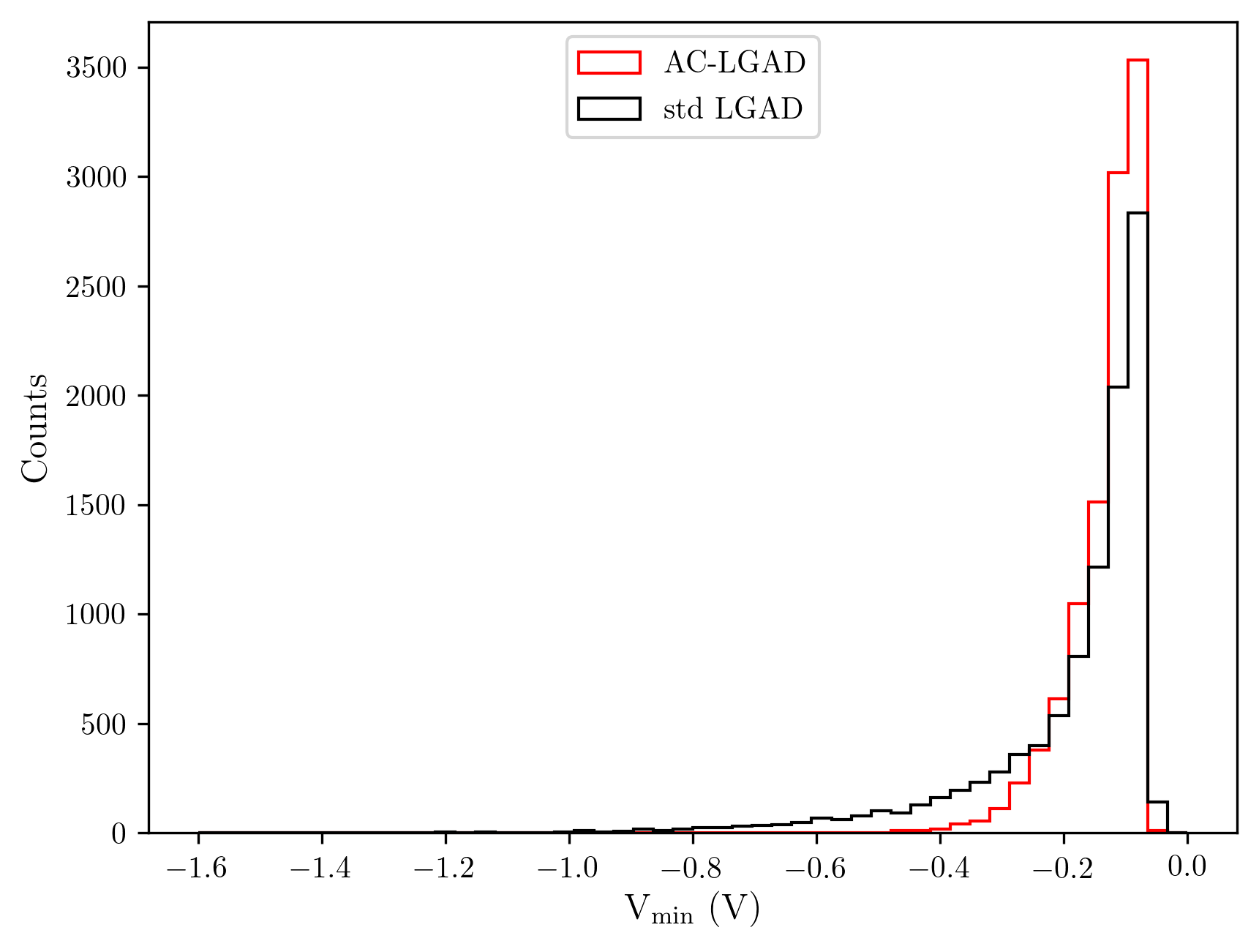}
\includegraphics[width=.43\textwidth]{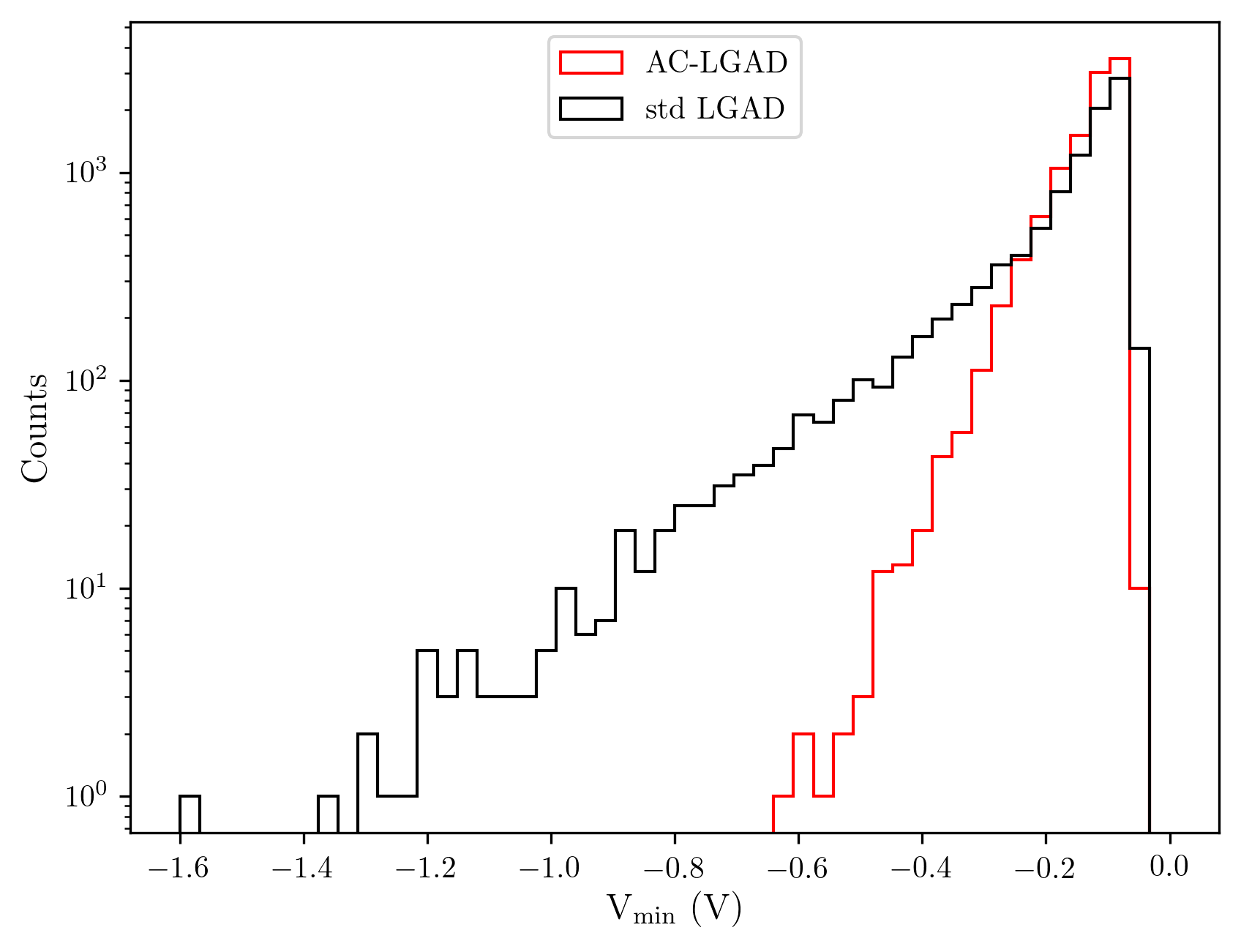}
\caption{\label{fig:LGADvsACLGAD} Pulse heights of about 10000 waveforms acquired from an LGAD (an AC-LGAD whose $n$++ was DC-connected to the TA) and a pixel of the AC-LGAD shown in figure~\ref{fig:sensors}. The bias voltage was set to $-$80 V while the trigger level at $-$10 mV in both cases. On the left the plot is in linear scale, on the right in logarithmic scale.}
\end{figure}

Using one of the nine  AC-coupled pads of the AC-LGAD shown in figure~\ref{fig:sensors}, waveforms generated by the same $^{90}$Sr source were acquired. The waveforms are primarily negative in voltages and the comparison of distributions of the peak amplitudes ($V_{\rm min}$) acquired by an LGAD (an AC-LGAD whose $n$++ was DC-connected to the TA) and by the AC-LGAD is shown in figure~\ref{fig:LGADvsACLGAD}. The $V_{\rm min}$ distributions peak at approximately the same values in the LGAD and in the AC-LGAD, which suggests that most of the charge is induced onto the AC-pixel: in the frequency range where the signals develop,  the impedance associated to the $C_{\rm AC}$ (20~pF, for this pad) is much lower than $R_{n}$.  The $V_{\rm min}$ distribution in the LGAD case shows a longer tail than the case of the AC-LGAD. This can be attributed to geometrical effects. In this AC-LGAD, the AC-pixel is much smaller than the whole active area of the LGAD (200 \textmu m $\times$  200 \textmu m as opposed to 1 mm $\times$  1 mm), therefore it collects less charge in case of 1) delta particles (that develop normally to the beta particle trajectory) and 2) beta particles from $^{90}$Sr that cross the silicon at an angle: these two events may induce charge in nearby pixels in the AC-LGAD, while the same charge is entirely collected in the large-area LGAD. 

 \subsection{Induced signal and cross-talk}


The induced signal produced in an AC-LGAD by different types of particle beams from radioactive sources is studied by simultaneously reading out three pixels in the 3 $\times$  3 array of the AC-LGAD with 200 \textmu m $\times$  200 \textmu m pixel size and 220 \textmu m pitch, see figure~\ref{fig:sensors} (a), using the above-mentioned TA multi-channel board: one channel is used for triggering the event, and the adjacent and next-to-adjacent channels, as defined in figure~\ref{fig:sensors} (b) and collectively referred to as {\it neighbouring channels}, are simultaneously read out. 

\begin{figure}
\centering 
\includegraphics[width=.6\textwidth]{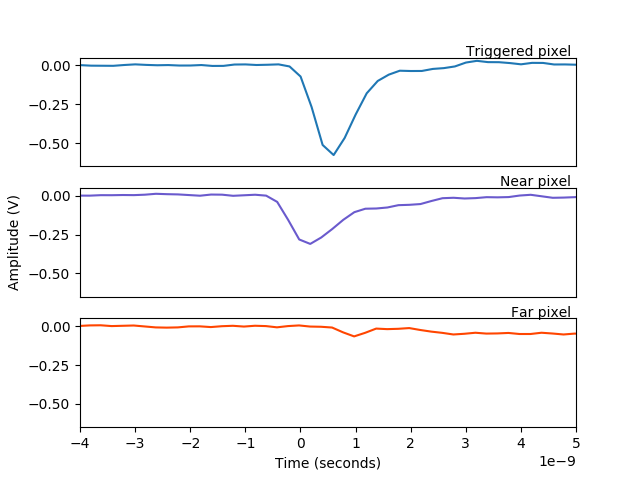}
\caption{\label{fig:sampleWf} Sample of waveforms recorded with a 1 GHz scope from a row of 3 channels in a 3 $\times$  3 array of AC-LGADs mounted to a multi-channel board. The waveforms are recorded for the triggering channel (Triggered pixel), its adjacent (Near pixel) and next-to-adjacent (Far pixel) channels, as sketched in figure~\ref{fig:sensors}. The signals are generated by beta radiation from a $^{90}$Sr source. A bias voltage of -80 V was applied and a trigger threshold of $-$100 mV was set on the scope on the triggering channel.
}
\end{figure}

\begin{figure}
\centering 
\includegraphics[width=.5\textwidth]{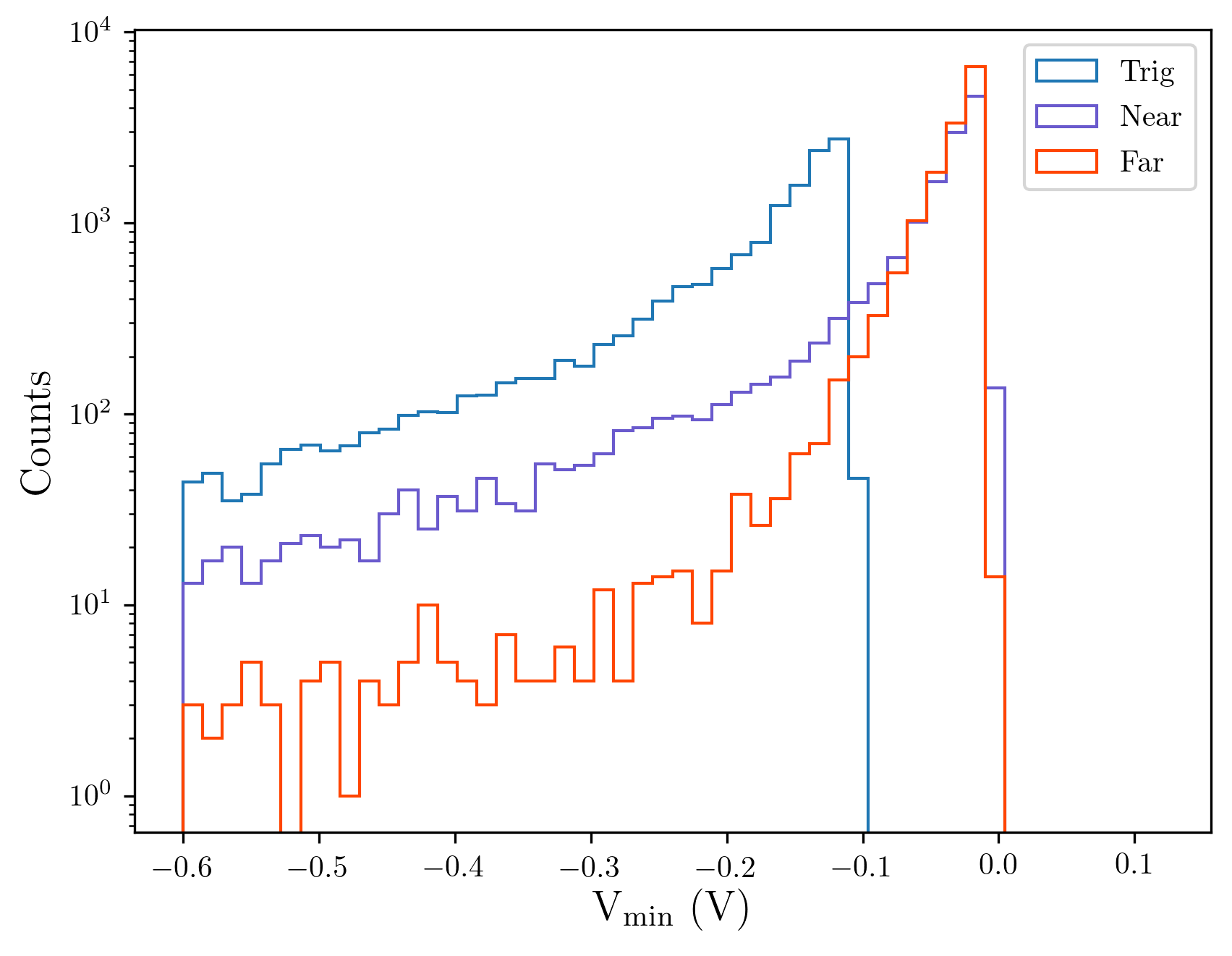}
\caption{\label{fig:minAmp}  Spectra of the amplitude  of signals produced by beta radiation from a $^{90}$Sr source for three channels in the 3 $\times$  3 array of AC-LGAD pixels in figure~\ref{fig:sensors}.  The blue line indicates the distribution of signals from the triggering pixel channel, while the purple and red lines refer to the pixel channels closer to and farther away from the triggering pixel, respectively. On the sensor a bias voltage of $-$80 V was applied and a trigger threshold of $-$100 mV was set on the scope on the triggering channel. The lower tails of the distributions are truncated at $-$0.6 V.
}
\end{figure}

\begin{figure}
\centering 
\hspace{-8mm}
\includegraphics[width=.5\textwidth]{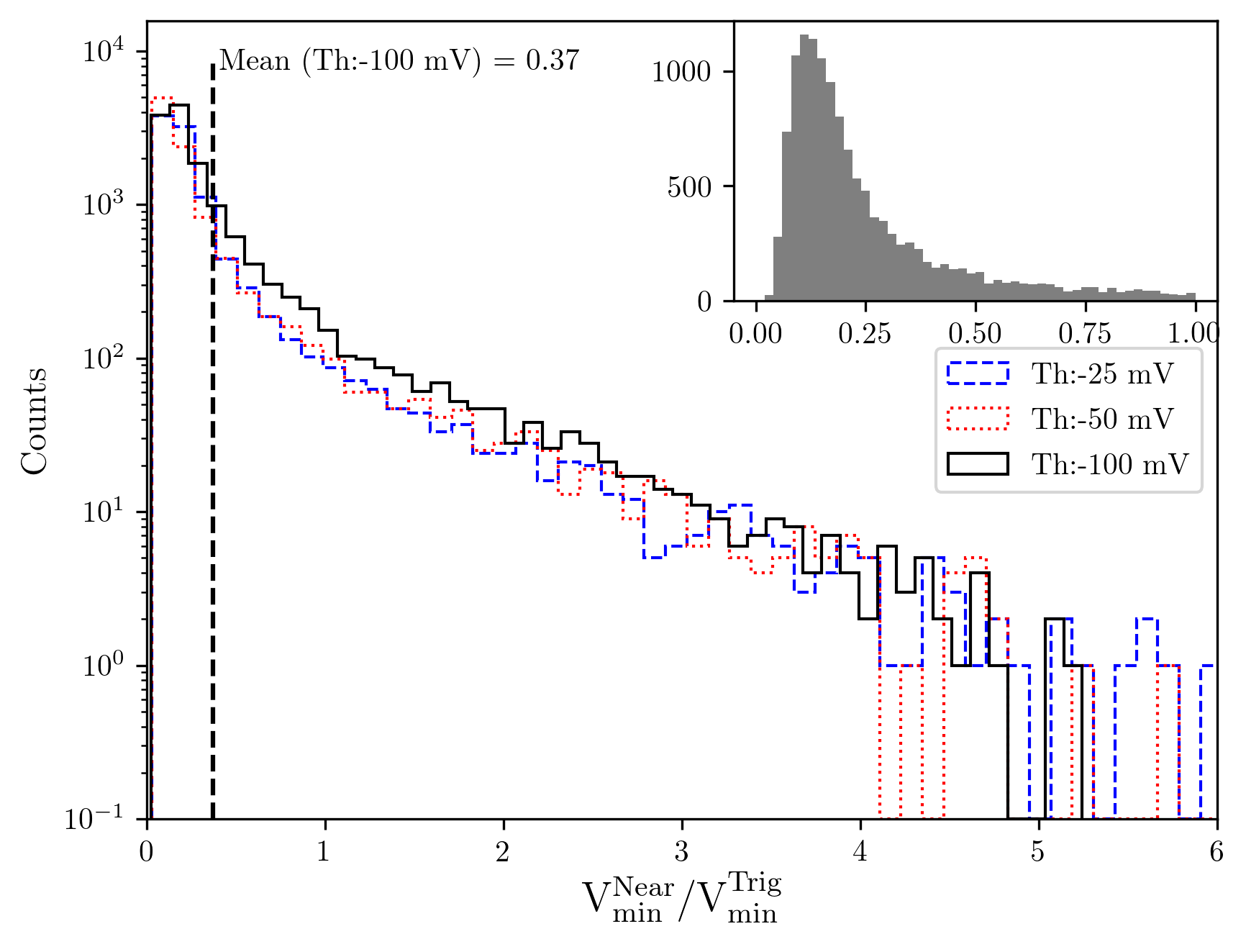}
\hspace{-2mm}
\includegraphics[width=.5\textwidth]{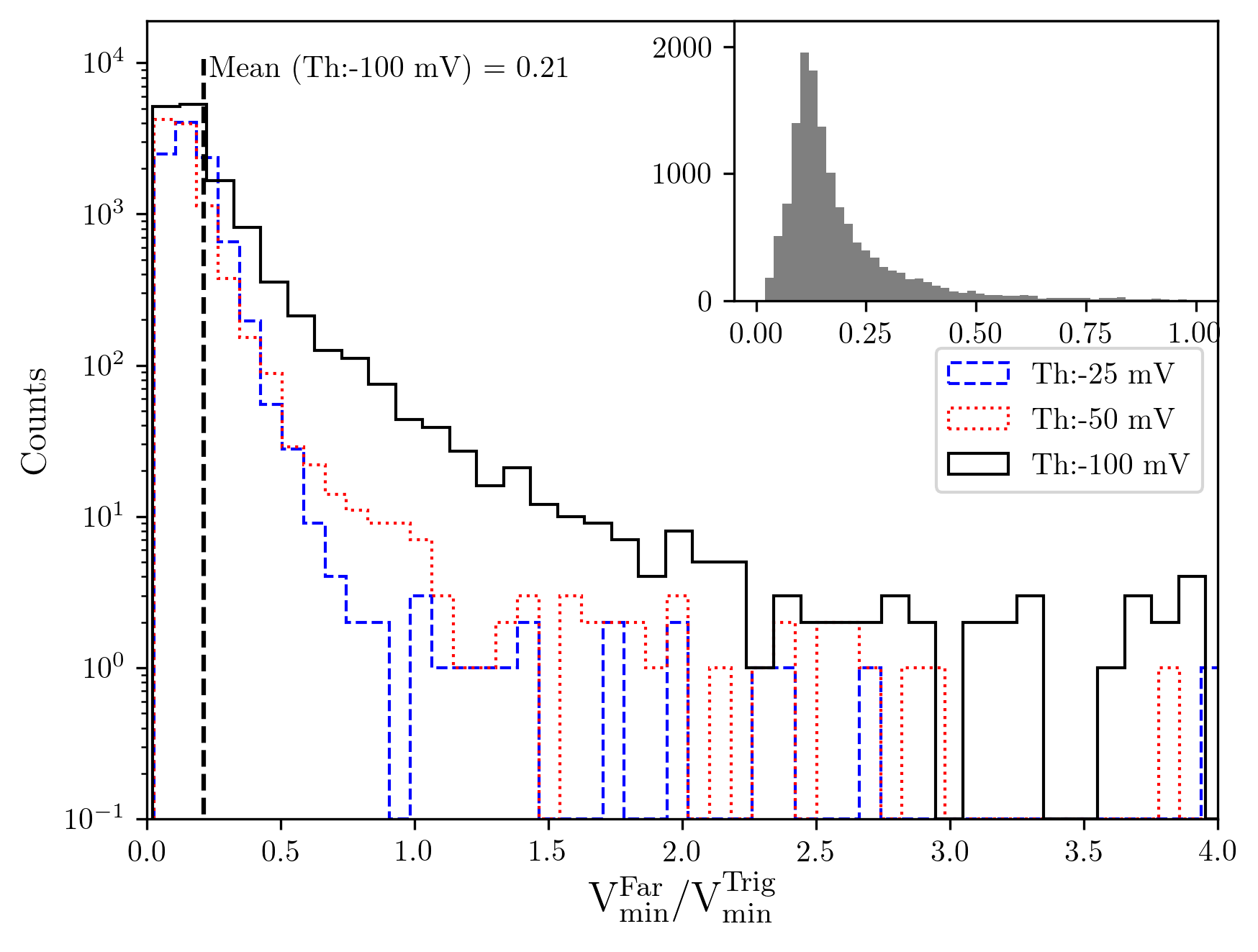}
\caption{\label{fig:ratioNT}  Distributions of the ratio between the  amplitude of signals produced in the pixel channel adjacent (left) and next-to-adjacent (right) to the triggering pixel and the one produced in the triggering pixel channel, for signals produced by a $^{90}$Sr source. The three different distributions in plots correspond to the three different trigger thresholds set on the scope in the triggering channel: $-$25, $-$50 and $-$100 mV. The black vertical dashed lines mark the mean values of the distributions. The inserts show with a linear scale on the vertical axes the distribution for the case with trigger threshold set at $-$100 mV. A bias voltage of $-$80 V was applied on the sensor.
}
\end{figure}

Figure~\ref{fig:sampleWf} shows a representative sample of signal waveforms recorded in the three channels with the 1 GHz scope, using beta particles from the $^{90}$Sr source impinging on the sensor. The signal waveforms  primarily extend to negative output voltages and show lower amplitudes the farther the pixel is located  from the triggering pixel, as expected.
%
%
The cross-talk between pixels is illustrated in figure~\ref{fig:minAmp}, which shows the comparison of the amplitude spectra  $V_{\rm min}$ for signals generated by beta radiation in the three nearby channels of the AC-LGAD device under study. For this test about 15,000 recorded events are used. 
Comparing the distributions for the triggering channel and its two neighbouring channels, we can see that the $V_{\min}$ distribution in the triggering channel starts from the trigger threshold of -100 mV and extends to lower $V_{\min}$ values, whereas neighbouring pixel channels exhibit spectra peaked at the lower amplitudes, because of the absence of trigger bias on those channels. Neighbouring pixels also feature progressively smaller tails in the $V_{\rm min}$ distribution the farther away the pixel is from the triggering pixel. This is compatible with signals of much lower induced charges for pixels far from the triggering pixel.  
%
%
The cross-talk between pixels is also studied as a function of the trigger threshold applied on the triggering channel and is quantified in figure~\ref{fig:ratioNT}, as the ratio of $V_{\rm min}$ of the neighbouring and triggering pixel channels for signal waveforms generated by the $^{90}$Sr beta source.
Given a trigger threshold on the triggering pixel channel of $-$100 mV, for the adjacent pixel the distribution of the ratios $V_{\rm min}^{\rm Near} / V_{\rm min}^{\rm Trig} $  has a mean value of 0.37 (uncertainty on the last digit), while for the next-to-adjacent pixel the distribution of the ratios $V_{\rm min}^{\rm Far} / V_{\rm min}^{\rm Trig} $  has a mean value of 0.21. For trigger values of $-$25 and $-$50 mV the mean values of ratios for adjacent (next-to adjacent) channels are in the range 0.32--0.33 (0.15--0.17).  The results show that, while the cross-talk between pixels is only weakly dependent on the trigger threshold, the tail of the $V_{\rm min}^{\rm Far} / V_{\rm min}^{\rm Trig} $ is significantly reduced when lower trigger thresholds are applied. This feature can be explained with the hypothesis that, with lower triggers, the triggered pixel records events that are further away from it and induce a fair amount of charge in the adjacent pixels, such that the $V_{\rm min}^{\rm Far} / V_{\rm min}^{\rm Trig} $ ratio has shorter tails. 


\begin{figure}
\centering 
\includegraphics[width=.45\textwidth]{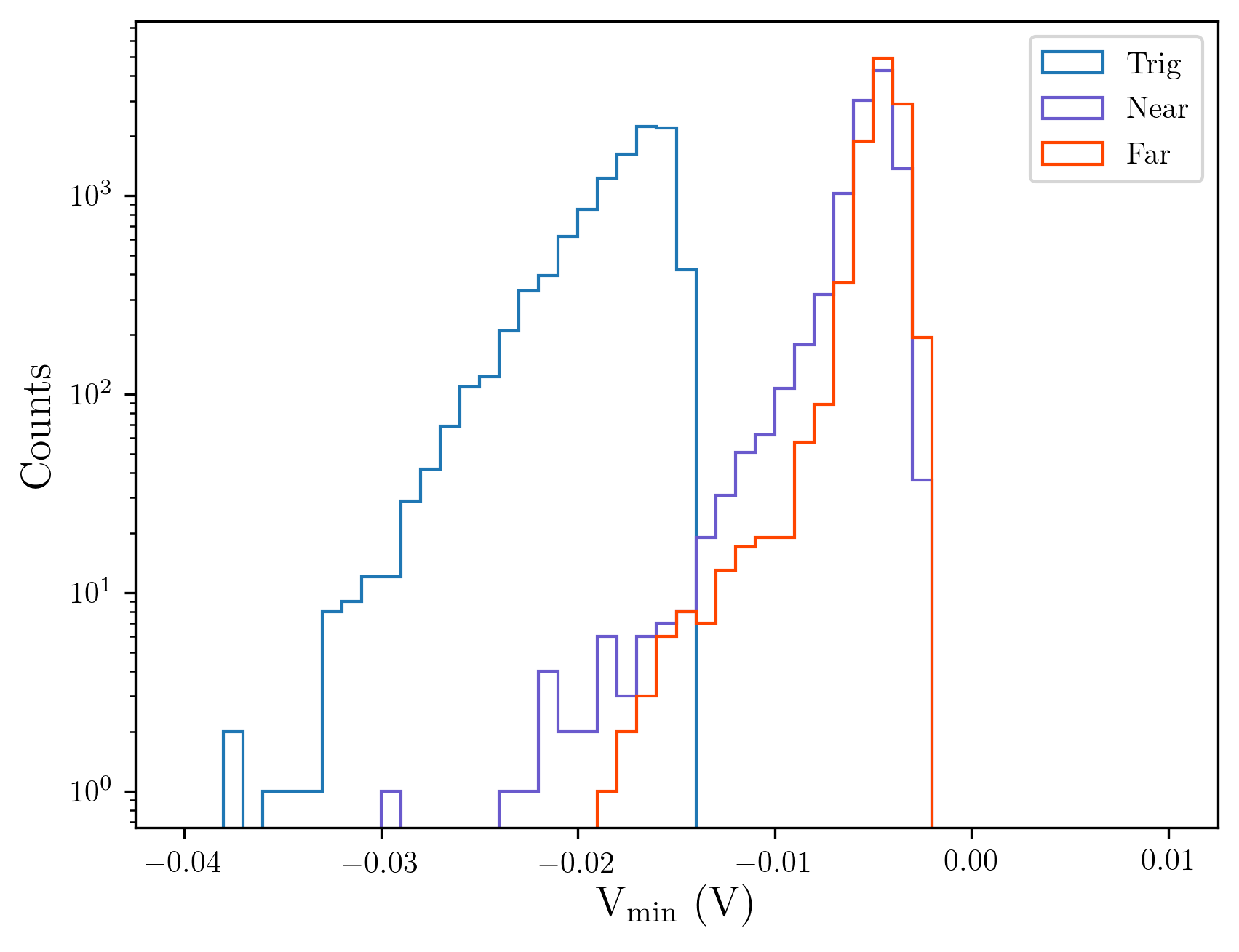}
\caption{\label{fig:minAmp-Fe_Am}  Distribution of the  amplitudes of the signal produced by X-rays generated from a $^{55}$Fe source 
 for the three channels in the 3 $\times$  3 array of AC-LGADs, as illustrated in figure~\ref{fig:sensors}.  The blue line indicates the distribution of signals from the triggering pixel channel, while the purple and red lines refer to the pixel channels closer to and farther away from the triggering pixel, respectively. 
}
\end{figure}

The above measurements were carried out using beta radiation from a $^{90}$Sr source and are representative of the sensor response to a particle beam close to minimum ionisation, which can penetrate the full substrate thickness. Tests were also conducted with X-rays that produce localised charge in the sensors, using $^{55}$Fe 
 radioactive sources that generate X-rays of about 6 keV. 
 Figure~\ref{fig:minAmp-Fe_Am} shows the $V_{\rm min}$ distributions for the three pixel channels under study in a sample of about 10,000 events. Comparing figure~\ref{fig:minAmp-Fe_Am} with figure~\ref{fig:minAmp}, we can see that the $V_{\rm min}$ distributions in the measurements with X-rays have shorter tails than those in the measurements with beta radiation. 

\begin{figure}
\centering 
\hspace{-8mm}
\includegraphics[width=.5\textwidth]{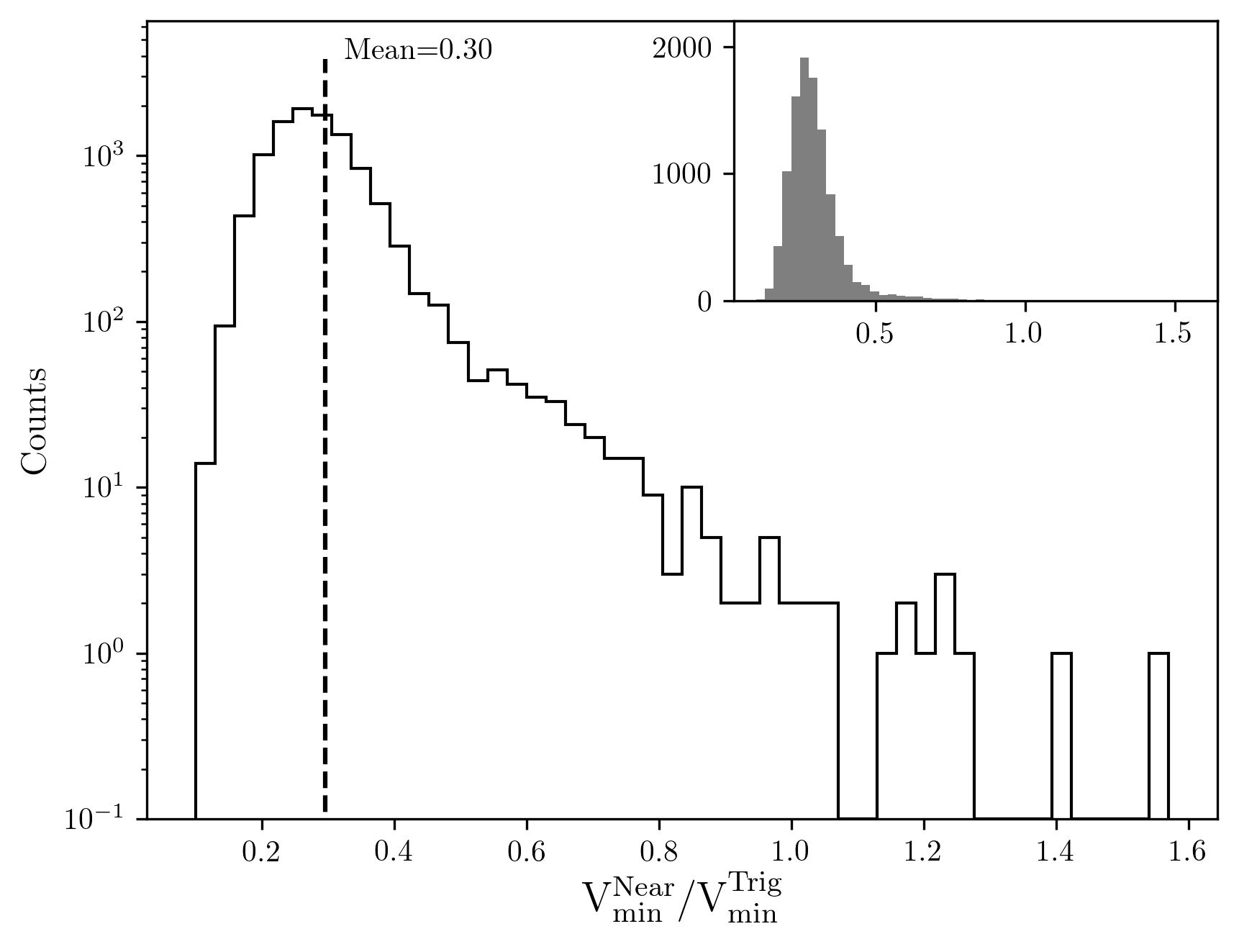}
\hspace{-2mm}
\includegraphics[width=.5\textwidth]{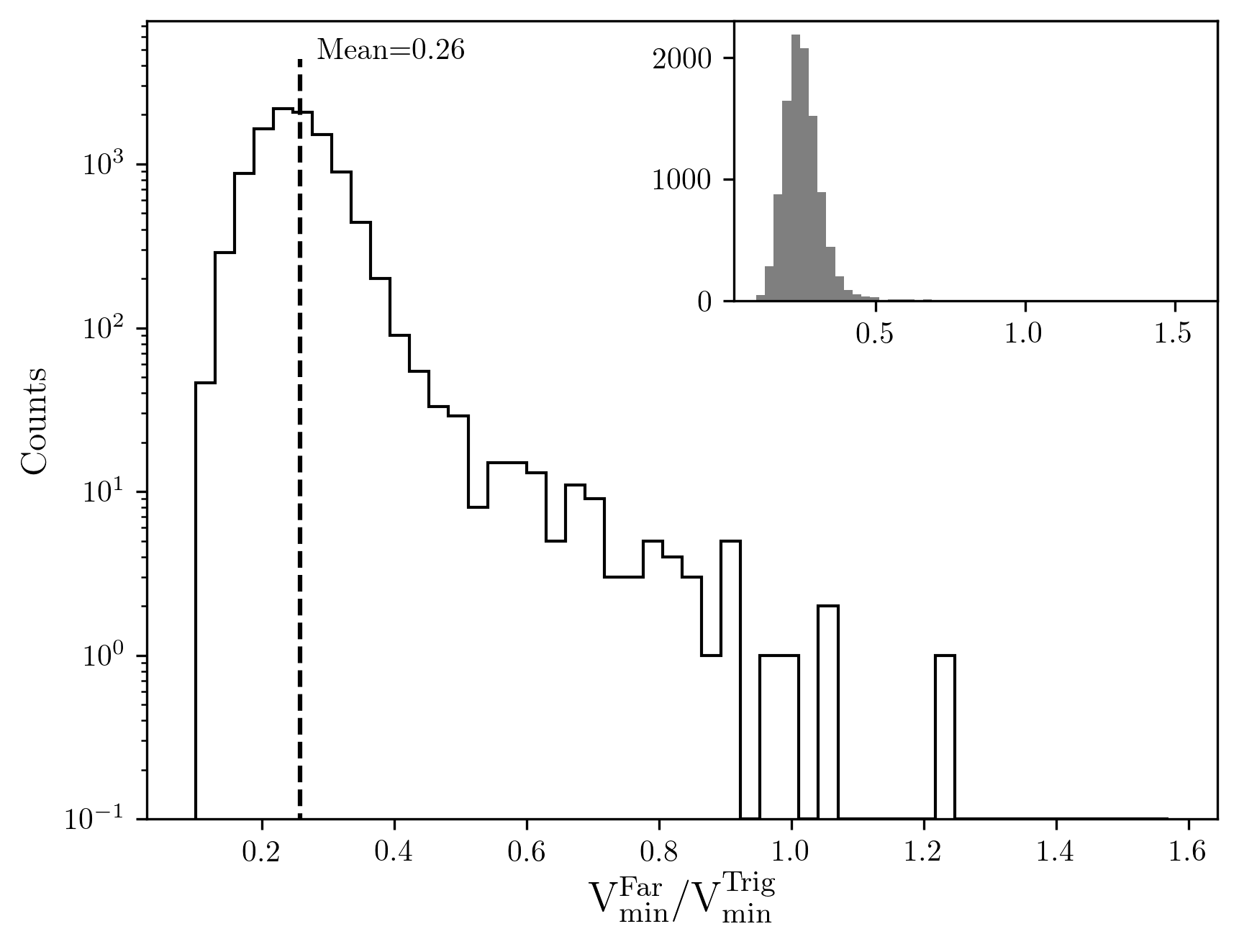}
\caption{\label{fig:ratioNT_FT_FE_AM}  Distribution of ratios of the signal amplitudes produced in the pixel channel adjacent (left) and next-to-adjacent (right) to the triggering pixel and those produced in the triggering pixel channel, for signals produced by X-rays generated from a $^{55}$Fe source. 
 The dashed vertical lines mark the mean values of the distributions. The inserts show the distributions with a linear scale on the vertical axes. 
}
\end{figure}

Similarly to the measurements with the $^{90}$Sr source, the ratios of the $V_{\rm min}$ between the neighbouring and triggering pixel channels for signal waveforms generated by X-rays from a $^{55}$Fe source  
are measured, see figure~\ref{fig:ratioNT_FT_FE_AM}.
The mean values of the ratios $V_{\rm min}^{\rm Near} / V_{\rm min}^{\rm Trig} $ and $V_{\rm min}^{\rm Far} / V_{\rm min}^{\rm Trig} $ are 
 0.30 and 0.26, respectively. The results show that the cross-talk between pixels 
 is similar to the one measured with beta radiation. However, it is worthy of attention the smaller ranges of the $V_{\rm min}^{\rm Near} / V_{\rm min}^{\rm Trig} $ and $V_{\rm min}^{\rm Far} / V_{\rm min}^{\rm Trig} $ distributions, which in the case of X-rays do not extend much further than 1. This observation is compatible with the hypothesis of localised charge creation in X-rays, as opposed to extended trajectories of beta particles passing through the sensor.

%
    %
 
 \subsection{Timing}



The time resolution $\sigma_t$ associated to the sensor jitter, driven by sensor noise, is calculated using the following formula (see ref.~\cite{spieler}, page 35):
\begin{align*}
\sigma_t=\frac{\sigma_{\rm noise}}{\frac{dS}{dt}}~,
\end{align*}
where $\sigma_{\rm noise}$ is the  r.m.s. voltage noise of the system, $S$ the mean signal  amplitude, and $\frac{dS}{dt}$ the slope of the signal ({\it slew rate}) as a function of the time $t$. 
Using as particle beam the beta radiation from a $^{90}$Sr source impinging on the 3 $\times$  3 array of AC-LGAD pixels as in figure~\ref{fig:sensors} with a bias voltage of $-$80 V, the $\sigma_{\rm noise}$ is estimated to be about 6 mV 
which from the calibration translates in the charge of about 2.5k electrons.
By calculating the slew rate in the time range between 10$\%$ and $90\%$ of the signal amplitude, the jitter $\sigma_t$ is estimated to reach approximately 20 ps.

\section{Conclusions and outlook}
A novel silicon device concept, the AC-LGAD, for fast-timing and high-granularity pixel detectors has been tested at BNL. Leveraging the experience on the fabrication of standard LGADs, wafers containing several AC-LGAD structures have been designed, fabricated and characterized. 
The specific design and fabrication process developed at BNL for AC-LGADs is outlined. The prototypes show good electrical characteristics, a gain value of 80, i.e., comparable to those of standard LGADs, and a timing resolution associated to the detector jitter that reaches approximately 20 ps. Further optimization of the fabrication process will increase the breakdown voltage by fine tuning the gain layer implantation parameters,  the resistance of the $n$+ layer  and the granularity of the pixelation. These improvements will allow to reach higher operational bias voltage, faster timing performance and finer spatial resolution.

\acknowledgments
The authors want to thank Enrico Rossi (BNL) for his help with the development and testing of programs for silicon sensor characterization, and Cinzia da Via (Manchester and Stony Brook Universities) for the useful discussions and support. The authors are indebted to Ronald Lipton and Artur Apresyan (FNAL) for the useful discussions and for providing the 16-channel board used for the measurements in this article.
This material is based upon work supported by the U.S. Department of Energy under grant DE-SC0012704.

\bibliographystyle{report}
\bibliography{biblio}{}
\end{document}